\documentclass[12pt]{article}
\usepackage{amsmath,amssymb}
\allowdisplaybreaks

\begin{document}
\baselineskip=16.0pt plus 0.2pt minus 0.1pt

\newcommand{\Tr}{\mathop{\textrm{Tr}}}
\newcommand{\Po}{\mathop{\textrm{P}}}
\newcommand{\Xw}{\mathop{{\cal X}}}
\newcommand{\ds}{\displaystyle}
\newcommand{\s}{\sigma}

\setcounter{page}{0}
\begin{flushright}
\parbox{40mm}{%
KUNS-1961 \\
{\tt hep-th/0504039} \\
April 2005}

\end{flushright}

\vfill

%%%%%%%%%%%%%%%%%%%%%%%%%%%%%%%%%%%%%%%%%%%%%%%%%%%%%
%% Title
%%%%%%%%%%%%%%%%%%%%%%%%%%%%%%%%%%%%%%%%%%%%%%%%%%%%
\begin{center}
{\Large{\bf 
BMN Operators from Wilson Loop
}}
\end{center}

\vfill

%%%%%%%%%%%%%%%%%%%%%%%%%%%%%%
%% author
%%%%%%%%%%%%%%%%%%%%%%%%%%%%%%
\begin{center}
{\sc Akitsugu Miwa}\footnote%
{E-mail: {\tt akitsugu@gauge.scphys.kyoto-u.ac.jp}}  \\[2em]
%$^1$
{\sl Department of Physics, Kyoto University, Kyoto 606-8502, Japan} \\

\end{center}

\vfill
%%%%%%%%%%%%%%%%%%%%%%%%%%%%%%%%%%
% Main
%%%%%%%%%%%%%%%%%%%%%%%%%%%%%%%%%%
\renewcommand{\thefootnote}{\arabic{footnote}}
\setcounter{footnote}{0}
\addtocounter{page}{1}
%%%%%%%%%%%%%%%%%%%%%%%%%%%%%%%%%%%%%%%%%%%%%%%%%%%%%%%%%%%%%%%%
%%%%%%%%%%%%%%%%%%%%%%%%%%%%%%%%%%%%%%%%%%%%%%%%%%%%%%%%%%%%%%%%
%%%%%%%%%%%%%%%%%%%
%% Abstract
%%%%%%%%%%%%%%%%%%%

\begin{center}
{\bf Abstract}
\end{center}

\begin{quote}

We show that the BMN operators arise from the expansion of the Wilson
loop in four-dimensional ${\cal N}=4$ super Yang-Mills theory.
The Wilson loop we consider is obtained from ``dimensional reduction''
of ten-dimensional ${\cal N}=1$ super Yang-Mills theory, and it
contains six scalar fields as well as the gauge field.
We expand the Wilson loop twice.
First we expand it in powers of the fluctuations around
a BPS loop configuration. Then we further expand each term in the
result of the first step in powers of the scalar field $Z$
associated with the BPS configuration.
We find that each operator in this expansion with large number of $Z$
is the BMN operator. The number of fluctuations corresponds to the
number of impurities, and the phase factor of each BMN
operator is supplied correctly.
We have to impose the locally supersymmetric 
condition on the loop for obtaining the
complete form of the BMN operators including the correction terms
with $\bar Z$.
Our observation suggests the correspondence between the Wilson loop
and the string field.

\end{quote}
\vfill
%%
%\baselineskip=\normalbaselineskip
%\renewcommand{\baselinestretch}{1.4}
%\setlength{\parskip}{0.3\baselineskip}
%%%%%%%%%%%%%%%%%%%%%%%%%%%%%%%%%%%%%%%%%%%%%%%%%%%%%%%%%%%%%%%%
%%%%%%%%%%%%%%%%%%%%%%%%%%%%%%%%%%%%%%%%%%%%%%%%%%%%%%%%%%%%%%%%

\newpage

%%%%%%%%%%%%%%%%%%%%%%%%%%
%%%%%%%%%%%%%%%%%%%%%%%%%%
\section{Introduction}
\label{introduction}
%%%%%%%%%%%%%%%%%%%%%%%%%%
%%%%%%%%%%%%%%%%%%%%%%%%%%
The AdS/CFT correspondence \cite{Maldacena:1997re,Gubser:1998bc,
Witten:1998qj,Aharony:1999ti,D'Hoker:2002aw}
is a conjecture that there is an equivalence
between type IIB superstring theory on $\textrm{AdS}_5 \times
\textrm{S}^5$ and four-dimensional ${\cal N} = 4$ SU($N$) super
Yang-Mills theory (SYM) on the boundary of $\textrm{AdS}_5$. This
correspondence was first proposed by Maldacena
\cite{Maldacena:1997re}, and a concrete holographic correspondence was
given by \cite{Gubser:1998bc,Witten:1998qj}. It has been intensively
studied as a realization of large $N$ duality between gauge theory and
string theory.
Until 2002, mainly the connections between supergravity modes
and BPS operators in SYM had been investigated. However, in paper
\cite{Berenstein:2002jq} the correspondence was extended to include
certain class of excited string modes and non-BPS operators. In the
string theory side, the authors of \cite{Berenstein:2002jq} considered
the string states on the pp wave background, which is the Penrose
limit of $\textrm{AdS}_5 \times \textrm{S}^5$. On the other hand in
the SYM side, they proposed certain class of operators with large
R-charge as the counterpart of the pp wave string states. This is now 
known as the BMN correspondence and these non-BPS operators are called
the BMN operators.
Although the BMN correspondence is still between the limited class of
string states and Yang-Mills operators, it is a strong support for the
full correspondence between these two theories.

On the other hand there is still another interesting connection
between SYM and string theory, or precisely, string field theory.
This is a correspondence between Wilson loop operator and string
field, and a correspondence between the equations of motion of the
two. String theory was originally investigated as a theory describing
the physics of hadrons, which is now known to be governed by QCD.
And there were a lot of works in which they tried to derive the area
law of the expectation value of the Wilson loop from its equation of
motion (loop equation)
\cite{Gervais:1978mp}.
The authors of these papers studied whether the loop equation can be
identified with the string equation of Nambu-Goto type.
Thus such an idea that there are some relations between the Wilson loop
and Nambu-Goto type string is quite an old one.
It is an interesting subject to reexamine this old idea from the view
point of the AdS/CFT correspondence.
In fact, there have been works which discuss the expectation value of
the Wilson loop in the context of the AdS/CFT correspondence and the
loop equation 
\cite{Rey:1998ik,Maldacena:1998im,Drukker:1999zq,Zarembo:2002an,Berenstein:1998ij,Gross:1998gk}.

The purpose of this paper is to show that the BMN operators are
found in the Wilson loop operator in SYM.%
\footnote{
In \cite{Imamura:2002xq} the BMN correspondence is reproduced by
considering the couplings among closed string states having large
angular momentum and open string states on D3-branes. This seems to be
a limited version of the coupling between a closed string and a Wilson
loop \cite{Rey:1998ik,Maldacena:1998im,Drukker:1999zq,Zarembo:2002an,Berenstein:1998ij,Gross:1998gk}. Thus the
observation in \cite{Imamura:2002xq} seems to be closely related to
ours.
}
Our Wilson loop is the one obtained from ``dimensional reduction'' of
ten-dimensional ${\cal N}=1$ SYM, and it consists of the six scalar
fields as well as the gauge field. Therefore, our loop is a ``loop in
ten-dimensional spacetime''.
We expand this Wilson loop operator twice. First we expand it
in powers of the loop fluctuation 
around the BPS configuration which is a point in four dimensions
and a straight-line in extra six dimensions.
Next we further expand each term of the first step in powers of the
scalar field $Z$ associated with the straight-line of the BPS
configuration.
We find that the generic operator in our double expansion 
is nothing but the BMN operator proposed in \cite{Berenstein:2002jq}.
The number of the impurities correspond to the number of the loop
fluctuations.
Furthermore we will see that even the correction terms
introduced and discussed in \cite{Parnachev:2002kk} can be reproduced
correctly. In particular, the correction term including the impurity
$\bar Z$ appears after we impose the locally supersymmetric condition
on the whole Wilson 
loop including the fluctuations. Our finding must be an important step
toward the understanding of the relation between the Wilson loop and
string field.  

This paper is organized as follows. In sec.\ \ref{BMNcor} we briefly review the
BMN correspondence. 
Sec.\ \ref{BMNfromW} is the main part of this paper. 
In subsection \ref{BMNsin} we perform the double
expansion we 
mentioned above. Then we show how the BMN operators proposed in
\cite{Berenstein:2002jq} with two or less impurities arise in the
expansion. In subsection \ref{BMNdou} we impose the locally
supersymmetric condition on the whole loop, and we show that the BMN
operators with correction terms given in \cite{Parnachev:2002kk} also
arise in the expansion.  Sec.\ \ref{conclusion} is devoted to the
conclusion and discussion. In appendix \ref{multi} we show that the
argument given in subsection \ref{BMNsin} can be generalized to the
case with $m$ impurities, and in appendix \ref{beyond} we give an
argument which does not need the saddle point approximation used in
sec.\ \ref{BMNfromW} and appendix \ref{multi}. 

%%%%%%%%%%%%%%%%%%%%%%%%%%%%%%%%%%%%%%%%%%%%%%%%
%%%%%%%%%%%%%%%%%%%%%%%%%%%%%%%%%%%%%%%%%%%%%%%%
\section{BMN correspondence}
\label{BMNcor}

Before discussing how the BMN operators arise in the Wilson loop, 
we shall recapitulate the BMN correspondence \cite{Berenstein:2002jq}.
This correspondence claims that the anomalous dimension of each BMN
operator coincides with the lightcone energy of the corresponding
string state propagating on the pp wave background.
The BMN operators are certain class of local operators in ${\cal N}=4$ 
SU($N$) SYM in four dimensions, and the pp wave
background is realized by 
taking the Penrose limit of $\textrm{AdS}_5 \times \textrm{S}^5$.
For this reason this correspondence is expected to be a restricted
version of the AdS/CFT correspondence.

We summarize in table \ref{BMN} the BMN correspondence proposed in
\cite{Berenstein:2002jq}.
\begin{table}
  \begin{center}
    \begin{tabular}{ccc}
      string states& & BMN operators \\ \hline
      $ |0\,;p^+ \rangle$ 
      & $\Longleftrightarrow$ 
      & $\Tr [Z^J]$ 
      \\
      $a^{M \dag}_0|0\,;p^+ \rangle $ 
      & $\Longleftrightarrow$
      & $\Tr [{\cal O}_M Z^J] $
      \\
      $a^{M_1 \dag}_{n_1} a^{M_2 \dag}_{n_2}|0\,;p^+\rangle$
      & $\Longleftrightarrow$ 
      & $\sum_{k_2=0}^{J} 
      \Tr[{\cal O}_{M_1} Z^{k_2} {\cal O}_{M_2}  Z^{J-k_2}]
      e^{2\pi i n_2 k_2/J}$
      \\[-2mm]
      $a^{M_1 \dag}_{n_1} \cdots a^{M_m \dag}_{n_m} |0\,;p^+ \rangle$
      &$\Longleftrightarrow$
      &\parbox{11cm}{
        \begin{align*}
          &\sum_{\sigma\in S_{m-1}}
          \sum\limits_{0\le k_{\sigma(2)}\le k_{\sigma(3)}\le
            \cdots\le k_{\sigma(m)}\le J}
          \hspace{-1cm}
          \Tr[{\cal O}_{M_1} Z^{k_{\sigma(2)}} {\cal O}_{M_{\sigma(2)}}
          Z^{k_{\sigma(3)}-k_{\sigma(2)}}\cdots
          \\[-1mm]
          &\qquad
          \cdots {\cal O}_{M_{\sigma(m-1)}}Z^{k_{\sigma(m)}-k_{\sigma(m-1)}}
          {\cal O}_{M_{\sigma(m)}} Z^{J-k_{\sigma(m)}}
          ] e^{2 \pi i \sum_{q=2}^m n_q k_q /J}
        \end{align*}
      }
      \\
      \hline
    \end{tabular}
  \end{center}
  \caption{
    The correspondence between the string states and the BMN operators
    proposed in \cite{Berenstein:2002jq}.}
  \label{BMN}
\end{table}
On the left hand side (LHS) of the table, $a^{M \dag}_n$ is the creation
operator of a string mode of level $n$ with $M=1,\ldots,8$ specifying
the transverse directions in the lightcone gauge.  
We have chosen a basis of the Fourier modes where $n>0$ ($n<0$)
corresponds to the left (right) mover.
Each string state on the LHS must satisfy the level matching condition
$\sum_{\ell=1}^m n_\ell =0$.
On the right hand side (RHS) of the table, $Z$ and ${\cal O}_M$
($M=1,\ldots, 8$) are defined by $Z=\Phi_5 + i \Phi_6$,
${\cal O}_{\mu} = D_\mu Z$ $(\mu = 1,\ldots,4)$ and
${\cal O}_{4+a} = \Phi_a$ $(a=1,\ldots,4)$ with $\Phi_i$ ($i=1,\ldots,6$)
being the six scalar fields in SYM.
Each operator on the RHS has a large number $J (\sim \sqrt{N})$ of $Z$ 
with some finite number of impurities ${\cal O}_M$ being inserted with
an appropriate phase factor. 
Summation is taken over the positions of impurities. 
In particular, the summation over $\sigma$ is with respect to the
permutations of $\{2,3,\ldots,m\}$.

In table \ref{BMN} we write only terms which contain impurities
whose classical conformal dimension $\Delta$
minus R-charge ${\cal J}$ is equal to $1$. 
In fact it has been known that these expressions need corrections
by terms which contain impurities with $\Delta - {\cal J} \geq 2$
\cite{Parnachev:2002kk}.
The complete forms of the large $J$ limit of the BMN operators
corresponding to the string states 
$a_{-n}^{\mu \dag} a_n^{\nu \dag} \big|0;p^+ \big\rangle$,
$a_{-n}^{4+a \dag} a_n^{\mu \dag}\big|0;p^+ \big\rangle$ and 
$a_{-n}^{4+a \dag} a_n^{4+b \dag} \big|0;p^+ \big\rangle$
are given respectively by
\begin{align}
  {\cal O}_{\mu \nu ,n}^J 
  &=
  \sum_{k=0}^J
  \Tr 
  \left[
    D_\mu Z Z^{k} D_\nu Z Z^{J-k}
  \right]
  e^{2 \pi i n k/J} 
  +
  \Tr\left[ D_{(\mu} D_{\nu)} Z Z^{J+1} \right] , \label{Omunun}\\
  {\cal O}_{4+a\,\mu,n}^J
  &=
  \sum_{k=0}^{J} 
  \Tr 
  \left[
    \Phi_a Z^{k} D_\mu Z Z^{J-k}  
  \right] 
  e^{ 2\pi i n k /J}
  +
  \Tr\left[ D_\mu \Phi_a Z^{J+1} \right], \label{Oimun}\\
  {\cal O}_{4+a \, 4+b,n}^J
  &=
  \sum_{k=0}^J 
  \Tr \left[ \Phi_a Z^{k} \Phi_b Z^{J-k} \right]
  e^{2 \pi i n k /J} 
  -
  \frac{1}{2}\delta_{ab} \Tr \left[ \bar{Z} Z^{J+1}  \right],
  \label{Oijn}
\end{align}
where $\bar{Z} = \Phi_5 - i\Phi_6$.
Other operators on the RHS of table \ref{BMN}
with three or more impurities need similar corrections.

%%%%%%%%%%%%%%%%%%%%%%%%%%%%%%%%%%%%%%%%%%%%%
\section{BMN operators from the expansion of Wilson loop}
\label{BMNfromW}

In this section we show that the BMN operators arise in the double
expansion of the Wilson loop operator. In subsection \ref{BMNsin} we
show how the impurities ${\cal O}_M$ ($M=1,\ldots,8$) with
$\Delta -{\cal J}=1$ appear with appropriate phase factors.
Next, in subsection \ref{BMNdou} we derive the correction terms with
$\Delta - {\cal J}=2$ in \eqref{Omunun} --- \eqref{Oijn}.

\subsection{Impurities with $\Delta  - {\cal J} =1$ 
and their phase factors}
\label{BMNsin}
Let us consider the Wilson loop operator consisting of the scalar
fields $\Phi_i$ as well as the gauge fields $A_\mu$
\cite{Rey:1998ik,Maldacena:1998im,Drukker:1999zq,Zarembo:2002an,Berenstein:1998ij,Gross:1998gk}:
\begin{align}
  W(C) = \Tr\!
  \left[
    {\rm P}
    \exp 
    \left(
      \int_0^t ds 
      \Big(
      A_{\mu}(x(s)) \dot{x}^\mu(s)
      +
      \Phi_i(x(s)) \dot{y}^i(s)
      \Big)
    \right)
  \right]
  = \Tr\big[ {\rm P} w_0^t(C) \big].   
  \label{Wilson loop}
\end{align}%
The existence of the scalar fields in the Wilson loop looks natural
if we recall that the four-dimensional ${\cal N}=4$ SYM is obtained
as a dimensional reduction of ten-dimensional ${\cal N}=1$ SYM.
The scalar fields are essential for reproducing the BMN operators and
also for deriving the loop equations (see sec.\ \ref{conclusion}).%
\footnote{
For the same reason, the fermionic extension of the Wilson loop is
also necessary. However, in this paper, we
concentrate only on the bosonic part.
}
As the loop $C$ we take $C=C_0 +\delta C$ with a BPS ``loop'' configuration
$C_0$ given by
\begin{align}
  x_{C_0}^\mu(s) = x^\mu, \qquad \dot{y}_{C_0}^i(s) =(0,0,0,0,1,i), 
\label{BPS}
\end{align}
and a small fluctuation 
$\delta C = \{ \delta x^\mu(s), \delta y^i(s)\}$  around it.
The Wilson loop \eqref{Wilson loop} is locally supersymmetric if the
condition 
\begin{equation}
\dot x(s)^2 - \dot y(s)^2=0
\label{BPScond}
\end{equation}
is satisfied and is $1/2$ BPS if the loop is a straight-line
\cite{Drukker:1999zq,Zarembo:2002an}.
Note that $W(C)$ is a functional of $x^\mu(s)$ and $\dot{y}^i(s)$, and
hence we take $x^\mu(s)$ and $\dot{y}^i(s)$ to be periodic functions
of $s$ with period $t$. This means that our Wilson ``loop'' is not
necessarily closed in the direction of $y^i$, but this dose not spoil
the gauge invariance of this operator. Therefore, we take
$\delta x^\mu (s)$ and $\delta \dot{y}^i(s)$ as periodic functions:
\begin{align}
  \delta x^\mu(s) = \sum_{n=-\infty}^\infty
  \delta x_n^\mu e^{2 \pi i n s/t},  \qquad
  \delta \dot{y}^i(s) = \sum_{n=-\infty}^\infty 
  \delta \dot{y}_n^i e^{2 \pi i n s/t}.
  \label{dy}
\end{align}
Note in particular that $\dot{y}^i$ and $\delta \dot{y}^i$ have zero
modes.
We Taylor-expand the Wilson loop $W(C) = W(C_0 + \delta C)$ around
$C_0$.%
\footnote{
Similar expansion is also considered in \cite{Dhar:2001ff}
in the context of the connection between the open Wilson line in
non-commutative Yang-Mills theory and the closed string field. 
In \cite{Berenstein:1998ij}, 
the expansion of the Wilson loop in terms of the
local operators is discussed in the context of the AdS/CFT and
the BMN correspondence. 
}
Since $W(C)$ is a functional of $x^\mu(s)$ and
$\dot{y}^i(s)$,
we have
\begin{align}
  W(C) 
  &= 
  W(C_0)
  +
  \int_0^t \! \! ds\, \delta x^\mu(s) 
  \frac{\delta W(C)}{\delta x^\mu(s)} \bigg|_{C=C_0} 
  +
  \int_0^t \! \! ds\,\delta \dot{y}^i(s)
  \frac{\delta W(C)}{\delta \dot{y}^i(s)} \bigg|_{C=C_0} \notag \\
  & \hspace{3cm}+ \frac{1}{2}
  \int_0^t \! \! ds_1 \int_0^t \! \! ds_2\, 
  \delta x^\mu(s_1) \delta x^\nu (s_2)
  \frac{\delta^2 W(C)}{\delta x^\mu(s_1) \delta x^\nu(s_2)} 
  \bigg|_{C=C_0} + \cdot \cdot \cdot  \notag \\
  &=
  W(C_0) 
  +
  \sum_{n} \delta x^\mu_n 
  \int_0^t \!\! ds 
  \frac{\delta W(C)}{\delta x^\mu(s)}\bigg|_{C=C_0}\hspace{-0.7cm}
  e^{2 \pi i n s/t }
  +
  \sum_{n} \delta \dot{y}^i_n
  \int_0^t \!\! ds \frac{\delta W(C)}{\delta \dot{y}^i(s)}
  \bigg|_{C=C_0}\hspace{-0.7cm}
  e^{2 \pi i n s/t } \notag \\
  & \hspace{3cm}
  + \frac{1}{2}
  \sum_{n_1,n_2} 
  \delta x^\mu_{n_1} \delta x^\nu_{n_2}
  \int_0^t \!\! ds_1 \int_0^t \!\! ds_2
  \frac{\delta^2 W(C)}
  {\delta x^{\mu}(s_1) \delta x^{\nu}(s_2)}\bigg|_{C=C_0}\hspace{-0.7cm}
  e^{2 \pi i (n_1 s_1 + n_2 s_2)/t} + \cdot \cdot \cdot.
\label{TaylorW(C)}
\end{align}  

Let us look at each term in this expansion in detail.
The first term $W(C_0)$ is given by
\begin{equation}
	W(C_0) =
	\Tr 
        \left[
          \exp \left(\int_0^t ds Z(x)\right)
        \right]
	= \sum_{J=0}^\infty \frac{t^J}{J!} 
        \Tr \left[Z^J\right]\!(x) ,
        \label{groundstate}
\end{equation}
where we have used that $\dot x^\mu(s)=0$ and
$\Phi_i(x(s))\dot{y}^i(s) = Z(x)$ on the path $C_0$.
Note that there appear the BMN operator $\Tr\left[Z^J\right]$
corresponding to the ground state of a pp wave string (see table 1).
Next we examine the terms which contain one functional derivative. 
For this, we use the following formula of functional derivatives:
\begin{align}
  \frac{\delta W(C)}{\delta x^\mu(s)} \Bigg|_{C=C_0} & = 
  \Tr 
  \left[ 
    \big(\, F_{\mu \nu}(x(s)) \dot{x}^\nu(s) 
    + 
    D_\mu \Phi_i(x(s)) \dot{y}^i(s) \, \big) \Po\big(w_s^{s+t}(C)\big)
  \right] \Big|_{C=C_0} \notag\\
  & =
  \Tr\left[D_\mu Z e^{tZ}\right]\!(x), 
\label{dWdx}
\\[2mm]
  \frac{\delta W(C)}{\delta \dot{y}^i(s)} \Bigg|_{C=C_0}
  &=
  \Tr
  \left[
    \Phi_i e^{tZ}\right]\!(x),
\label{dWdy}
\end{align}
where $w_s^{s'}(C)$ represents, as in eq.\ \eqref{Wilson loop}, the
Wilson line along $C$ from $s$ to $s'$ without trace.
Then we have
\begin{align}
  \sum_{n} \delta x^\mu_n 
  \int_0^t \!\! ds 
  \frac{\delta W(C)}{\delta x^\mu(s)}\bigg|_{C=C_0}\hspace{-0.7cm}
  e^{2 \pi i n s/t}
	&=
	\sum_{n} \delta x^\mu_n 
  \int_0^t \!\! ds \Tr\! 
  \left[ 
    D_\mu Z e^{tZ}  \right]\!(x) 
  e^{2 \pi i n s/t} \notag \\
	&=
	\delta x^\mu_0 \sum_{J=0}^\infty \frac{t^{J+1}}{J!}
	\Tr\left[ D_\mu Z Z^J  \right]\!(x).
        \label{SUGRAX}
\end{align} 
The $s$-integration yields the Kronecker delta $t \delta_{n,0}$,
which picks up only the zero mode, namely the SUGRA mode, from the
summation over $n$. The SUGRA modes in the other directions are
supplied from functional derivatives with respect to $\dot{y}^i(s)$:
\begin{equation}
  \sum_{n} \delta \dot{y}^i_n 
  \int_0^t \!\! ds \frac{\delta W(C)}{\delta \dot{y}^i(s)}
  \bigg|_{C=C_0}\hspace{-0.7cm}
  e^{2 \pi i n s/t}
  =
  \delta \dot{y}^i_0 \sum_{J=0}^\infty \frac{t^{J+1}}{J!}
  \Tr\left[ \Phi_i  Z^J  \right]\!(x) .
  \label{SUGRAY}
\end{equation}
The SUGRA modes correspond to $i=1,\ldots,4$, while the meaning of the
$i=5, 6$ terms is not clear at the present stage.
Here we simply ignore the $i=5,6$ terms. However, we shall see in
subsection \ref{BMNdou} that these terms play important roles in
reproducing the correction terms.
From \eqref{SUGRAX} and \eqref{SUGRAY}, we expect the following
correspondence between the functional derivatives and the oscillation
modes of the pp wave string (see table \ref{BMN}):
\begin{align}
  \int_0^t ds \frac{\delta}{\delta x^\mu(s)} 
  \longrightarrow
  a_0^{\mu \dag}, \qquad 
  \int_0^t ds \frac{\delta}{\delta \dot{y}^a(s)}
  \longrightarrow
  a_0^{4+a \dag}. 
\end{align}

Next we shall study whether higher string modes can be derived in the
same manner. For this purpose we consider the terms which contain two
functional derivatives in the expansion \eqref{TaylorW(C)} of the
Wilson loop.
Here we use the dilute gas approximation and assume that the
parameters $s_1$ and $s_2$ of the two functional derivatives are not
close to each other.
In particular, we omit the contribution from the terms proportional to
$\delta(s_1-s_2)$ and $\partial_{s_1} \delta(s_1 - s_2)$. We will
examine these terms in the next subsection.
With this approximation, the following replacement formulas similar to
\eqref{dWdx} and \eqref{dWdy} hold:
\begin{align}
  \frac{\delta}{\delta x^\mu (s)} \to D_\mu Z(x(s)), \qquad
  \frac{\delta}{ \delta \dot{y}^i(s)} \to \Phi_i (x(s)).   
\end{align}
Thus we must consider terms of the following form:
\begin{align}
  &  \int_0^t ds_1 \int_{s_1}^{s_1 + t} \!\!\!ds_2
  \Tr \left[ 
    {\cal O}_{M_1}(x) 
    w_{s_1}^{s_2}(C_0)
    {\cal O}_{M_2}(x) 
    w_{s_2}^{s_1 + t}(C_0)
  \right]
  e^{2 \pi i(n_1 s_1 + n_2 s_2)/t}\notag\\
  &=\int_0^t ds_1 \int_{s_1}^{s_1+t} \!\!\!ds_2 
  \sum_{J=0}^\infty
  \sum_{k_2=0}^J 
  \frac{1}{(J-k_2)!k_2!}(s_2-s_1)^{k_2} (s_1 + t -s_2)^{J-k_2} \notag \\
  & \hspace{6cm}\times \Tr
  \left[
    {\cal O}_{M_1}  Z^{k_2} {\cal O}_{M_2} Z^{J-k_2} 
  \right]\!(x)\, e^{2 \pi i(n_1 s_1 + n_2 s_2)/t}.
  \label{operator2} 
\end{align}
By adopting the new integration variables
$(\tilde{s}_1,\tilde{s}_2)=(s_1,s_2-s_1)/t$,
the integration over $ \tilde{s}_1$ can be performed to yield the
Kronecker delta $t \delta_{n_1 + n_2,0}$, and \eqref{operator2} is
further rewritten into
\begin{equation}
  \delta_{n_1+n_2,0}
  \sum_{J=0}^\infty t^{J+2} 
  \sum_{k_2=0}^J
  \Tr
  \left[
    {\cal O}_{M_1}
    Z^{k_2} 
    {\cal O}_{M_2}
    Z^{J-k_2} 
  \right]\!(x) 
  F_2(n_2,k_2,J),
\label{operator2sumJ} 
\end{equation}
where we have defined
\begin{equation}
    F_2(n_2,k_2,J)
    \equiv
    \frac{1}{(J-k_2) !k_2!}
  \int_0^1 d \tilde s_2 
  \tilde s_2^{k_2} (1- \tilde s_2)^{J-k_2}
  e^{2 \pi i n_2 \tilde s_2}.
  \label{F2}
\end{equation}
In the large $J$ limit, the integration over $\tilde s_2$ can be evaluated
using the saddle point approximation to give\footnote{
For the validity of the saddle point approximation, it is necessary
that both $k_2/J$ and $1-(k_2/J)$ are of $O(1)$ as can be seen from
the coefficients of the non-Gaussian terms of the fluctuation around
the saddle point.
Since the saddle point is at $\tilde s_2=k_2/J$, this condition is
consistent with the dilute gas assumption.
} 
\begin{align}
  F_2(n_2,k_2,J) 
  \sim \frac{1}{J J!}
  \exp\left( \frac{2 \pi i n_2 k_2}{J} \right),
\label{saddles2}
\end{align}
where we have used Stirling's formula, i.e., 
$n! \sim \sqrt{2 \pi  n}\,n^n e^{-n}$ ($n\gg 1$).
The saddle point of the integration \eqref{F2} is at
$\tilde s_2 =k_2/J$.
Finally we can evaluate the operators in the $J$-summation
of \eqref{operator2sumJ} for large $J$ limit as  
\begin{align}
  t^{J+2}
  \frac{\delta_{n_1+n_2,0}}{J J!}\sum_{k_2 = 0}^J 
  \Tr 
  \left[
    {\cal O}_{M_1} Z^{k_2} {\cal O}_{M_2} Z^{J-k_2}
  \right]\!(x)
  e^{2 \pi i n_2 k_2/J}.
\label{BMN2}
\end{align}
This is nothing but the BMN operator with two impurities given in
table \ref{BMN}.
As seen from the above derivation, the expression \eqref{BMN2} may be
incorrect in the region where $k_2/J$ is close to $0$ or $1$, namely,
when the two impurities ${\cal O}_{M_1}$ and ${\cal O}_{M_2}$ are
close to each other.
However, we shall show in appendix \ref{beyond} that \eqref{BMN2} is
in fact valid for all $k_2$ so long as $J$ is large.

%%%%%%%%%%%%%%%%%%%%%%%%%%%%%%%%%%%%%%%%%%%%%%%%%%%%%%%%%
%%%%%%%%%%%%%%%%%%%%%%%%%%%%%%%%%%%%%%%%%%%%%%%%%%%%%%%%%
\subsection{Complete forms of the large $J$ BMN operators} 
\label{BMNdou}

In this subsection we show that the BMN operators with the correction
terms, \eqref{Omunun} --- \eqref{Oijn}, arise from the double 
expansion of the Wilson loop \eqref{TaylorW(C)}.
The origins of the correction terms are i) two functional derivatives
acting at the same point on the loop which we neglected in the
previous subsection, and ii) the functional derivatives
$\delta/\delta \dot y^i$ with $i=5,6$ together with the
locally supersymmetric constraint on the fluctuation of the loop which
we newly impose in this subsection.
 
First, two functional derivatives acting on $W(C)$ is given, including 
the delta function terms, by
\begin{align}
  \frac{\delta^2  W(C)}{\delta x^\mu(s_1) \delta x^\nu(s_2)}
 & \Biggr|_{C=C_0} =
  \theta(s_2-s_1)\Tr 
  \left[
    w_0^{s_1}(C_0) D_\mu Z(x) w_{s_1}^{s_2}(C_0) D_\nu Z(x) w_{s_2}^t(C_0) 
  \right] \notag \\[-4.5mm]
  & \hspace{2cm}+
  \theta(s_1-s_2)
  \Tr 
  \left[
    w_0^{s_2}(C_0) D_\nu Z(x) w_{s_2}^{s_1}(C_0) D_\mu Z(x) w_{s_1}^t(C_0) 
  \right] \notag \\
  &\hspace{2cm}+ \delta(s_1 - s_2)
  \Tr 
  \left[
    w_0^{s_1}(C_0) D_{(\mu} D_{\nu)} Z(x) w_{s_1}^t(C_0) 
  \right] \notag \\[-2mm]
  &\hspace{-1.5cm}+ \frac{1}{2}\partial_{s_2} \delta(s_1 - s_2)
  \Tr
  \left[
    w_0^{s_1}(C_0)F_{\mu \nu}(x(s_1))w_{s_1}^t(C_0)
    +
    w_0^{s_2}(C_0)F_{\mu \nu}(x(s_2))w_{s_2}^t(C_0)
  \right],
  \label{ddW/dxdx} \\[2mm]
  \frac{\delta^2  W(C)}{\delta \dot y^i(s_1) \delta x^\mu(s_2)}
 & \Biggr|_{C=C_0} =
  \theta(s_2-s_1)\Tr 
  \left[
    w_0^{s_1}(C_0) \Phi_i(x) w_{s_1}^{s_2}(C_0) D_\mu Z(x) w_{s_2}^t(C_0) 
  \right] \notag \\[-4.5mm] 
  & \hspace{2cm}+
  \theta(s_1-s_2) 
  \Tr 
  \left[
    w_0^{s_2}(C_0) D_\mu Z(x) w_{s_2}^{s_1}(C_0) \Phi_i (x) w_{s_1}^t(C_0) 
  \right] \notag \\
  &\hspace{2cm}+ \delta(s_1 - s_2)
  \Tr 
  \left[
    w_0^{s_1}(C_0) D_\mu \Phi_i(x) w_{s_1}^t(C_0) 
  \right],
  \label{ddW/dydx} \\[2mm]
  \frac{\delta^2  W(C)}{\delta \dot y^i(s_1) \delta \dot y^j(s_2)}
 & \Biggr|_{C=C_0} =
  \theta(s_2-s_1)\Tr 
  \left[
    w_0^{s_1}(C_0) \Phi_i (x) w_{s_1}^{s_2}(C_0) \Phi_j (x) w_{s_2}^t(C_0) 
  \right] \notag \\[-4.5mm]
  & \hspace{2cm}+
  \theta(s_1-s_2)
  \Tr 
  \left[
    w_0^{s_2}(C_0) \Phi_j(x) w_{s_2}^{s_1}(C_0) \Phi_i(x) w_{s_1}^t(C_0) 
  \right],
  \label{ddW/dydy} 
 \end{align}
with $\theta(s)=1$ ($= 0$) for $s>0$ ($s<0$), and 
$D_{(\mu}D_{\nu)}\equiv\left(D_\mu D_\nu+D_\nu D_\mu\right)/2$.
We have used the Bianchi identity of $F_{\mu\nu}$ in obtaining
\eqref{ddW/dxdx} which is manifestly symmetric under the exchange
$(s_1,\mu)\leftrightarrow(s_2,\nu)$. 
In the previous subsection, we omitted terms proportional to
$\delta(s_1-s_2)$ and $\delta'(s_1-s_2)$.
Using \eqref{ddW/dxdx} --- \eqref{ddW/dydy}, the terms with two
functional derivatives in \eqref{TaylorW(C)} are reduced to
\begin{align}
   &\int_0^t ds_1 \int_0^t ds_2
   \delta x^\mu(s_1) \delta x^\nu(s_2)
   \frac{\delta^2 W(C)}
   {\delta x^{\mu} (s_1) \delta x^{\nu} (s_2)}\biggr|_{C=C_0}
   \notag \\
   &\hspace{1cm}=
    \sum_J 
  \sum_{n=-\infty}^\infty t^{J+2}\delta x^\mu_{-n} \delta x^\nu_{n}
  \bigg\{
    \sum_{k=0}^{J} \Tr 
    \left[ D_\mu Z Z^{k} D_\nu Z Z^{J-k} \right] F_2(n,k,J)
    \notag \\[-3mm]
    & \hspace{6cm}
    + \frac{1}{(J+1)!} \Tr\left[ D_{(\mu} D_{\nu)} Z Z^{J+1} \right] 
    - \frac{2 \pi i n}{(J+2)!} 
    \Tr \left[ F_{\mu \nu} Z^{J+2}\right]
  \bigg\} \notag \\[-3mm]
  &\hspace{1cm}\sim   \sum_J \frac{1}{J J!}
  \sum_{n=-\infty}^\infty t^{J+2} \delta x^\mu_{-n} \delta x^\nu_{n}
  {\cal O}_{\mu \nu ,n}^J
,  \label{ints^2ddW/dxdx}\\[3mm]
   &\int_0^t ds_1 \int_0^t ds_2
   \delta \dot y^a(s_1) \delta x^\mu(s_2)
   \frac{\delta^2 W(C)}
   {\delta \dot y^a (s_1) \delta x^{\mu} (s_2)}\biggr|_{C=C_0}  \notag \\
  &\hspace{1cm}
  =\sum_J \sum_{n=-\infty}^\infty  t^{J+2} \delta \dot y^a_{-n} \delta x^\mu_{n}
  \bigg\{
  \sum_{k=0}^{J}
  \Tr 
  \left[
    \Phi_a Z^{k} D_\mu Z Z^{J-k}   \right]F_2(n,k,J) 
  + \frac{1}{(J+1)!} \Tr\left[ D_\mu \Phi_a Z^{J+1} \right]
  \bigg\} \notag \\
   &\hspace{1cm}\sim
  \sum_J \frac{1}{J J!} 
  \sum_{n=-\infty}^\infty t^{J+2}\delta \dot y^a_{-n} \delta x^\mu_{n}
  {\cal O}_{4+a\,\mu ,n}^J
  , \label{ints^2ddW/dydx} \\[3mm]
  &\hspace{0cm}\int_0^t\!\! ds_1 \!\! \int_0^t\!\! ds_2
  \delta \dot y^a(s_1) \delta \dot y^b(s_2)
   \frac{\delta^2 W(C)}
   {\delta \dot y^a (s_1) \delta \dot y^b (s_2)}\biggr|_{C=C_0} 
   \hspace{-0.6cm}=\sum_J\! \!\sum_{n=-\infty}^\infty \!\!
   t^{J+2} \delta \dot y^a_{-n}\delta \dot y^b_{n} \sum_{k=0}^J
  \Tr\! \left[ \Phi_a Z^{k} \Phi_b Z^{J-k} \right]
  \!F_2(n,k,J) \notag \\
   &\hspace{4cm}\sim
      \sum_J \frac{1}{JJ!} 
   \sum_{n=-\infty}^\infty  t^{J+2}
   \delta \dot y^a_{-n}\delta \dot y^b_{n} 
   \sum_{k=0}^J 
   \Tr \left[ \Phi_a Z^{k} \Phi_b Z^{J-k} \right]
   e^{2 \pi i n k /J}. \label{ints^2ddW/dydy}
\end{align}
In the last line of each equation we have taken the large $J$ limit
using \eqref{saddles2}. 
Furthermore, in \eqref{ints^2ddW/dxdx}, we have neglected the last
term containing $F_{\mu \nu}$ in the second expression, 
because this term is multiplied by an additional factor $1/J$ compared
to the other two terms.
In \eqref{ints^2ddW/dydx} and \eqref{ints^2ddW/dydy},
the indices $a$ and $b$ run from $1$ to $4$, and we have omitted
the fluctuations $\delta \dot y^5$ and $\delta \dot y^6$. We will see
later that $\delta \dot y^5$ and $\delta \dot y^6$ contribute to the
terms with more powers of the transverse fluctuations.
In \eqref{ints^2ddW/dxdx} and \eqref{ints^2ddW/dydx} we have obtained
the correct operators ${\cal O}_{\mu \nu,n}^J$ \eqref{Omunun} and
${\cal O}_{4+a\,\mu,n}^J$ \eqref{Oimun}, respectively.
On the other hand, the operator in the last expression of
\eqref{ints^2ddW/dydy} differs from ${\cal O}_{4+a\,4+b,n}^J$
\eqref{Oijn} in that the former lacks the term
$- \delta_{ab}\Tr[\bar{Z}Z^{J+1}]/2$. 
This is because the four directions $\delta \dot y^a$
($a=1,\ldots,4$)  
do not directly correspond to the transverse directions of the  pp
wave string modes in the lightcone gauge.
In the following, we shall 
show that the correct choice of the four coordinates can be
found by imposing the locally supersymmetric condition \eqref{BPScond}
on the whole loop including the fluctuations.\footnote{
In \cite{Maldacena:1998im},
this locally supersymmetric condition is imposed on the Wilson loop
operator from the start.
}
 
Before proceeding to discuss the locally supersymmetric condition, we
shall give some comments on our double expansion.  
Introducing the new coordinates $z(s) \equiv (y^5(s) + i y^6(s))/2$ and 
$\bar{z}(s) \equiv (y^5(s) - i y^6(s))/2$, we have
$\sum_{i=5,6}\Phi_i \dot y^i=Z\dot{\bar z} + \bar Z\dot z$,
and $C_0$ \eqref{BPS} is rewritten as
\begin{equation}
  x_{C_0}^\mu(s) = x^\mu, 
  \qquad \dot y_{C_0}^a(s) = \dot z_{C_0}(s) = 0, 
  \qquad \dot{\bar{z}}_{C_0}(s) = 1.
\end{equation}
Using the reparametrization invariance, we can fix
$
\dot{\bar{z}}(s)=\dot{\bar{z}}_{C_0}(s)+\delta \dot{\bar{z}}(s)
$
equal to its value of $C_0$:
\begin{equation}
   \dot{\bar{z}}(s) = 1.
\label{z}
\end{equation}
In other words, we can always gauge away the small fluctuation $\delta
\dot{\bar{z}}(s)$, i.e., $\delta \dot{\bar{z}}(s) = 0$.
Using this gauge, our double expansion is reduced to the following two
steps. First  we expand around the straight-line $C_0$
with respect to the nine coordinates $\delta x^\mu(s)$, 
$\delta \dot y^a(s)$ and $ \delta \dot{z}(s)$. 
Next we expand this straight-line itself with respect to $Z$.

Now let us impose the locally supersymmetric condition \eqref{BPScond}
on the whole loop including the fluctuation:    
\begin{equation}
  \big(\dot{x}_{C_0}^\mu(s) +\delta \dot x^\mu(s) \big)^2 
  -
  \big(\dot{y}_{C_0}^i(s) + \delta \dot y^i(s)  \big)^2
  =
  \big(\delta \dot x^\mu(s) \big)^2
  -  4 \dot{\bar{z}}(s) \delta \dot z(s)
  - \big(\delta \dot y^a(s) \big)^2
  =0.\label{constraint}
\end{equation}
Using the gauge condition \eqref{z} we can solve this constraint for
$\delta \dot{z}(s)$:
\begin{align}
  \delta \dot{z}(s)
  &= \frac{1}{4}\left(
    \big(\delta \dot x^\mu(s) \big)^2
    -
    \big(\delta \dot y^a(s) \big)^2 
  \right). \label{dz}
\end{align} 
This constraint does supply the missing correction term for
\eqref{ints^2ddW/dydy} mentioned above.
To see this we reexamine the terms containing one
functional derivative with respect to $\dot y^5(s)$ and  $\dot y^6(s)$
in \eqref{TaylorW(C)}: 
\begin{align}
  \int_0^t ds
  \left(
  \delta \dot y^5(s) \frac{\delta W(C)}{\delta \dot y^5(s)}
  +
  \delta \dot y^6(s) \frac{\delta W(C)}{\delta \dot y^6(s)}
  \right)\biggr|_{C=C_0}
  =
  \int_0^t ds
    \delta \dot z(s) \frac{\delta W(C)}{\delta \dot z(s)}
    \biggr|_{C=C_0} ,
    \label{dW/d5+dW/d6todW/dz}
\end{align}
where we have used the gauge condition $\delta \dot{\bar{z}}(s) = 0$.
Using 
\begin{align}
  \frac{\delta W(C)}{\delta \dot z(s)} \biggr|_{C=C_0}
  =
  \sum_{J=0}^\infty \frac{t^J}{J!}
  \Tr \left[ \bar{Z} Z^J\right](x), \qquad
  \label{dW/dzdW/dbz}
\end{align}
and the locally supersymmetric condition \eqref{dz}, we obtain
\begin{align}
  \int_0^t ds \delta \dot z(s) \frac{\delta W(C)}{\delta \dot z(s)}
  \biggr|_{C=C_0}
  &=
   \frac{1}{4} 
   \sum_{J=0}^\infty \frac{1}{J!} 
  \sum_{n = -\infty}^\infty
  \left(
    (2 \pi n)^2 t^{J-1}\delta x^\mu_{-n} \delta x^\mu_n
    -
    t^{J+1}\delta \dot y^a_{-n} \delta \dot y^a_n
  \right)
 \Tr [ \bar{Z}
  Z^J](x) ,
  \label{intsdW/dz}
\end{align}
which are quadratic in the fluctuations and should be added to
\eqref{ints^2ddW/dxdx} and \eqref{ints^2ddW/dydy}.
For \eqref{ints^2ddW/dxdx}, the $(\delta x)^2$ term of
\eqref{intsdW/dz} adds $\delta_{\mu \nu} (2 \pi n)^2 
\Tr{\bar{Z} Z^{J+3}}/2(J+3)!$ to the second line of \eqref{ints^2ddW/dxdx}
(here we have adjusted the conformal dimension and the R-charge of
the added operator).
However, the added operator has an extra factor $1/J^2$
compared to  ${\cal O}_{\mu \nu,n}^J$ and hence we can neglect it. 
On the other hand, we should add $-\delta_{ab} \Tr{\bar{Z}
  Z^{J+1}}/2(J+1)!$ to the last line of \eqref{ints^2ddW/dydy}.
This term is exactly what is needed in order to have the correct BMN
operator ${\cal O}_{4+a\,4+b,n}^J$.
Thus the locally supersymmetric condition \eqref{dz} extracts the
correct fluctuations corresponding to the transverse directions of the
pp wave string. In the same manner as we have shown in 
\eqref{dW/d5+dW/d6todW/dz} --- \eqref{intsdW/dz}, the terms containing
functional derivatives with respect to $\dot y^5(s)$ and $\dot y^6(s)$
in \eqref{ddW/dydx} and \eqref{ddW/dydy} contribute to the string
states with three or more oscillators.
 
In this section we have shown that the BMN operators corresponding
to the string states with two or less excited modes arise in the
Taylor expansion of the Wilson loop operator \eqref{TaylorW(C)}. 
In subsection \ref{BMNsin}, neglecting the contribution from two
functional derivatives acting at the same point, we derived the
correspondence between the fluctuation modes of the string and those
of the Wilson loop. In appendix \ref{multi} we generalize
this argument to the BMN operators with an arbitrary number of
impurities. This observation strongly suggests that we can reinterpret
the BMN correspondence, and moreover the full AdS/CFT correspondence, as
the relation between a string and a Wilson loop. Further discussion
including the relation between the equations of motion of these
dynamical variables will be given in the next concluding section. 
In subsection \ref{BMNdou} we have shown that the correction terms in
\eqref{Omunun} --- \eqref{Oijn} are also found in the expansion
\eqref{TaylorW(C)}. The correction terms in \eqref{Omunun} and
\eqref{Oimun} arise from two functional derivatives acting at the same
point which we neglected in subsection \ref{BMNsin}.
However, as for \eqref{Oijn}, the existence of the
correction term is the result of the mismatch between the two sets of
four directions \{$\delta y_n^a$\} and \{$a_n^{4+a \dag}$\}. 
Imposing the locally supersymmetric condition \eqref{dz} we have
extracted the correct four directions and reproduced the correction
term in \eqref{Oijn} from the expansion of the Wilson loop.  

Finally we give the large $J$ part of the expansion of $W(C)$:
\begin{align}
  W(C) &\sim
  \sum_J\frac{t^J}{J!}
  \Bigg\{
  {\cal O}_{\textrm{ground}}^J
    +
    t
    \left(\delta x^\mu_0   {\cal O}_{\mu,0}^J
    +
    \delta \dot y^a_0 {\cal O}_{4+a,0}^J \right)\notag \\
    &\hspace{1cm}+
    \frac{t}{2!}\left(
    \sum_n\delta x^\mu_{-n} \delta x^\nu_n {\cal O}_{\mu \nu ,n}^{J-1}
    +
    \sum_n\delta \dot y^a_{-n} \delta x^\mu_n {\cal O}_{4+a\,\mu ,n}^{J-1}
    +
    \sum_n\delta \dot y^a_{-n} \delta \dot y^b_n {\cal O}_{4+a\,4+b ,n}^{J-1}
    \right)+
    \cdots
  \Bigg\}.
\end{align}
with ${\cal O}_{\textrm{ground}}^J = \Tr [Z^J]$, ${\cal O}_{\mu,0}^J
= \Tr[D_\mu Z Z^J]$ and ${\cal O}_{4+a,0}^J = \Tr[\Phi_a Z^J]$.
The dots represents terms with three or more fluctuations.

%%%%%%%%%%%%%%%%%%%%%%%%%%%%%%%%%%%%%%%%%%%%%
%%%%%%%%%%%%%%%%%%%%%%%%%%%%%%%%%%%%%%%%%%%%%
\section{Conclusion and discussion}
\label{conclusion}
%%%%%%%%%%%%%%%%%%%%%%%%%%%%%%%%%%%%%%%%%%%%%
%%%%%%%%%%%%%%%%%%%%%%%%%%%%%%%%%%%%%%%%%%%%%

\begin{table}
  \begin{center}
  \begin{tabular}{ccc}
    SYM side
    & &
    string side \\ \hline
    $W(C_0)$ 
    & $\longrightarrow$ &
    $|0;\,p^+\rangle$ \\[3mm]
    $\ds\int_0^t ds \frac{\delta}{\delta x^\mu(s)} e^{2 \pi i n s/t}$
    & $\longrightarrow$ &
    $a_n^{\mu \dag}$ \\[5mm]
    $\ds\int_0^t ds \frac{\delta}{\delta \dot{y}^a(s)} e^{2 \pi i n s/t}$
    & $\longrightarrow$ &
    $a_n^{4+a \dag}$ \\[3mm]  \hline
  \end{tabular}
  \end{center}
  \caption{The map from the functional derivatives 
    to the excited string modes. The locally supersymmetric condition
    is necessary to give the correction terms.}
  \label{summary}
\end{table}% 
In this paper we clarified how the BMN operators, including the
correction terms, are embedded in the Wilson loop operator in
four-dimensional ${\cal N} = 4$ SYM.  
First we expanded the Wilson loop in powers of the fluctuations around
the BPS configuration \eqref{BPS}. Then we further expanded each term
in this series in powers of the scalar field $Z$.
We saw that the operators with large number of $Z$ in this expansion
are nothing but the BMN operators. 
The nontrivial phase factors in the
BMN operators appear in a natural way from the mode functions 
$e^{2\pi i ns/t}$ for the fluctuations of the loop, \eqref{dy}.
For the generic configurations of the functional derivatives,
the number of the impurities in the BMN operator coincides with the
power of the fluctuations, which is equal to the number of functional
derivatives operating the Wilson loop.
Using this identification we can reinterpret the
BMN correspondence as the relation between the fluctuation modes of the
Wilson loop and those of the string.
This correspondence is summarized in table \ref{summary}.
We have also shown that, in order to derive the correction terms
of the large $J$ BMN operators, naive identification of the
directions $(\mu,a)$ such as the one given in table \ref{summary}
without any constraint is not correct. This is because there is some
mismatch between the directions of \{$\delta \dot y_n^a$\} and
\{$a_n^{4+a \dag}$\}. We must 
extract the correct four directions from \{$\delta \dot y^i$\}
corresponding to the transverse directions of the pp wave string modes
in the lightcone gauge. This extraction was realized by imposing the
locally supersymmetric condition \eqref{dz}.

The correspondence in table \ref{summary} connects only the narrow
sectors in both of the two theories, SYM and string theory.
In the string side the spacetime geometry
is the pp wave which is the Penrose limit of
$\textrm{AdS}_5\times \textrm{S}^5$.
On the other hand, in the SYM side the corresponding local operators
in the expansion of the Wilson loop are limited to those containing
a large number of scalar fields $Z$.
However, we expect that there is a full correspondence between the
Wilson loop operator in four-dimensional ${\cal N}=4$ SYM and the type
IIB superstring field on the
$\textrm{AdS}_5\times\textrm{S}^5$ background. Furthermore we expect
that there is a correspondence between the equations of motion which
these dynamical variables obey. In the following, we shall discuss
these points more concretely.

First consider the expansion of the Wilson loop in powers of
the fluctuations:
\begin{equation}
  W(C) = \sum {\cal O}_{\{M,n\}}(x) \delta X^{M_1}_{n_1} 
  \delta X^{M_2}_{n_2} \cdots \delta X^{M_m}_{n_m},
\label{W(C)}
\end{equation}
where $\delta X$ is one of the fluctuations $\{ \delta x^\mu , \delta
y^i \}$, and the operators ${\cal O}_{\{M,n\}}(x)$ 
are written in terms of local fields in SYM. On the other hand, the
string field $\Psi[X=X_0+\delta X]$ which is a functional of the
fluctuation $\delta X(\sigma)$ around the zero-mode $X_0$
can be expanded in terms of spacetime component fields as
\begin{equation}
  \Psi[X(\sigma)] = \sum \psi_{\{M,n\}}(X_0) 
  \bigl\langle \delta X\bigr|
   a^{M_1 \dag}_{n_1}a^{M_2 \dag}_{n_2} \cdots a^{M_m \dag}_{n_m}
  \bigl|\textrm{ground state} \bigr\rangle,
\label{Psi[X]}
\end{equation}
where $a^{M \dag}_n$ are the creation operators in the Hilbert space
of the first quantized string, and $\psi_{\{M,n\}}(X_0)$ are the
component fields.
Note that $\bigl\langle \delta X\bigr|
a^{M_1 \dag}_{n_1}a^{M_2 \dag}_{n_2} \cdots a^{M_m \dag}_{n_m}
\bigl|\textrm{ground state} \bigr\rangle$
on the RHS of \eqref{Psi[X]}, namely, the wave function of the string
state in the $\delta X_n^M$-diagonal representation, is roughly equal
to  $\delta X^{M_1}_{n_1} \delta X^{M_2}_{n_2} \cdots
\delta X^{M_m}_{n_m}$ on the RHS of \eqref{W(C)} up to 
a Gaussian like function of $\delta X_n^M$.
Now we mean, by the correspondence between the Wilson loop and the
string field, that there exists a complete map between the operators
 and the component fields:
\begin{equation}
{\cal O}_{\{M,n\}}(x)  
\,\Longleftrightarrow\,
\psi_{\{M,n\}}(X_0).
\end{equation}
Our analysis sec.\ \ref{BMNfromW} supports this correspondence in the
pp wave limit, and we expect that the full $\textrm{AdS}_5 \times
\textrm{S}^5$ version of the correspondence does exist.

Next let us discuss the equations of motion.
The equation of motion of string field on the curved geometry $G_{MN}$
should be given by
\begin{align}
\int\!d\s\! \left\{
    -G^{MN}(X)\frac{\delta^2}{\delta X^M\delta X^N}
    + 
    G_{MN}(X){X^M}' {X^N}' 
  \right\}\!(\sigma)
  \Psi[X]+\left(\mbox{interaction terms}\right)=0,  
  \label{Hamiltonian}
\end{align} 
and we expect that the Wilson loop $W(C)$ (by multiplied the Gaussian
like function mentioned above) satisfies the same equation as 
\eqref{Hamiltonian}.
In fact, it has been known that the Wilson loop operator is subject to
the loop equation which resembles \eqref{Hamiltonian}
\cite{Gervais:1978mp}.\footnote{
An interesting connection between the IIB
matrix model and the string field theory is also discussed in
\cite{Fukuma:1997en}.   
}
The starting point of the loop equation is the following formula for
the second functional derivative of $W(C)$:
\begin{align}
  \frac{\delta^2}{\delta X^M (s_1) \delta  X^N(s_2)} W(C)
  &= {{X}^P}'(s_1) {{X}^Q}'(s_2)
  \Tr \Bigg[
  \Po \Big(
  F_{MP}\big(x(s_1)\big) F_{NQ}\big(x(s_2)\big)
   w_0^t(C)
  \Big)
  \Bigg] \notag\\
&\qquad
+\big(\mbox{terms with }\delta(s_1-s_2)
\mbox{ or } \delta'(s_1-s_2)\big).  \label{loop equation 00}
\end{align}
We want to derive from \eqref{loop equation 00} an equation for $W(C)$
which is equivalent to \eqref{Hamiltonian}.
This would be possible if the equation like
\begin{align}
G^{MN}(X(s))F_{MP}(X(s)) F_{NQ}(X(s))\sim G_{PQ}(X(s)),
\end{align}
is realized in some sense for the $\textrm{AdS}_5 \times \textrm{S}^5$
metric (a modification due to the Gaussian like function is
necessary).
The last delta function terms of \eqref{loop equation 00} is expected
to contribute the interaction terms.
Further investigation on this subject is necessary.

\section*{Acknowledgment}

I would like to thank Y.\ Aisaka, M.\ Asano, M.\ Fukuma, 
K.\ Hashimoto, H.\ Hata, T.\ Higaki, H.\ Kawai, Y.\ Kimura, T.\
Morita, S.\ Moriyama, K.\ Murakami, M.\ Sato, S.\ Seki, H.\ Shimada
and S.\ Sugimoto for valuable discussions and suggestions. I also
would like to thank G.\ Georgiou for giving me stimulating comments
after the fist version of this paper appeared on the arXiv. 
I am grateful to H.\ Hata for careful reading of the manuscript. 
This work is supported in part by the Grant-in-Aid for the 21st Century COE 
``Center for Diversity and Universality in Physics'' from the Ministry 
of Education, Culture, Sports, Science and Technology (MEXT) of Japan.

%%%%%%%%%
%%%%%%%%%
\appendix
%%%%%%%%%
%%%%%%%%%

%%%%%%%%%%%%%%%%%%%%%%%%%%%%%%
%%%%%%%%%%%%%%%%%%%%%%%%%%%%%%
\section{Multi-impurities} \label{multi}
%%%%%%%%%%%%%%%%%%%%%%%%%%%%%%
%%%%%%%%%%%%%%%%%%%%%%%%%%%%%%
In this appendix we present the calculation of the general terms in
the Taylor expansion of the Wilson loop operator.
Neglecting the contribution from the coincident $s$ configurations,
the term containing $m$ functional derivatives is given by the sum of
the following type of operators over the permutations of
$\{{\cal O}_{M_2},{\cal O}_{M_3},\cdots,{\cal O}_{M_m}\}$:
\begin{align}
  &\int_0^t ds_1 \int_{s_1}^{s_1+t} \!\!\!ds_2\cdots
  \int_{s_{m-1}}^{s_1+t} ds_m
  \Tr
  \big[
    {\cal O}_{M_1}(x) w_{s_1}^{s_2}(C_0)
    {\cal O}_{M_2}(x) w_{s_2}^{s_3}(C_0)
    \cdots \notag \\
    & \hspace{7cm} \cdots
    w_{s_{m-1}}^{s_m}(C_0) 
    {\cal O}_{M_m}(x)
    w_{s_m}^{s_1+t}(C_0)
  \big] e^{2 \pi i \sum_{q=1}^m n_q s_q/t}.  
\end{align}
Repeating the same steps as in sec.\ \ref{BMNfromW}, we can rewrite it as
\begin{align}
  \delta_{\sum_{q=1}^m n_q,0}
  \sum_{J=0}^\infty 
  \sum_{0\leq k_2 \leq k_3 \leq \cdots \leq k_m \leq J}
  &\!\!t^{J+m} \Tr
  \left[
    {\cal O}_{M_1}
    Z^{k_2}
    {\cal O}_{M_2}
    Z^{k_3-k_2}
    \cdots
    Z^{k_m-k_{m-1}}
    {\cal O}_{M_m}
    Z^{J - k_m}
  \right]\!(x)
  \notag\\
  &\hspace{3.5cm} \times 
  F_m\big(\{n_q\},\{k_q\},J\big), 
  \label{multiBMN}
\end{align}
where $F_m$ is defined by
\begin{align}
  F_m\big(\{n_q\},\{k_q\},J\big) &\equiv
  \frac{1}
  {k_2!(k_3-k_2)! \cdots (k_m-k_{m-1})!(J-k_m)!} \notag \\
  &  \times
  \left(
  \prod_{q=2}^{m}\int_0^{1-\sum_{r=2}^{q-1}\tilde{s}_r}
  d \tilde{s}_q \tilde{s}_q^{k_q-k_{q-1}} 
  \exp
  \left( 
    2 \pi i
    \sum_{r=q}^m n_r \tilde{s}_q
  \right)
  \right)
  \left(1- \sum_{r=2}^m \tilde{s}_r\right)^{J - k_m} ,
\label{F_m}
\end{align}
with $k_1\equiv 0$ and $\sum_{r=2}^1 \tilde s_r =0$.
The new integration variables with tilde are
$\tilde{s}_q=\big(s_q-s_{q-1}\big)/t$ ($q = 2,\ldots,m$).
We carry out the $\tilde s$-integrations in the large
$k_q-k_{q-1}$ and $J-k_m$  limit by using the saddle point
approximation as we did in sec.\ \ref{BMNfromW}. 
The saddle point is at  
$\tilde{s}_q = (k_q - k_{q-1})/J$, and we obtain 
\begin{align}
  F_m\big(\{n_q\},\{k_q\},J\big) 
  & \sim
  \frac{1}{J^{m-1}J!} 
  \exp\left( \frac{2 \pi i \sum_{q=2}^m n_q k_q}{J}\right).
  \label{saddlesm}
\end{align}
As we noted in sec.\ \ref{BMNfromW}, this expression is valid only in
the region 
of $k_q$ where none of $(k_q - k_{q-1})/J$ nor $1-k_m/J$
is close to zero, in other words, in the region where the impurities
are well separated from each other.
We show in appendix \ref{beyond}
that the expression \eqref{saddlesm} is in fact
justifiable also when two of $k_q$ are close together.
Using \eqref{saddlesm}, we find that the large $J$ limit of the
operators in the $J$-summation of \eqref{multiBMN} is given by
\begin{align}
  &\frac{t^{J+m}\delta_{\sum_{q=1}^m n_q,0}}{(J+m-1)!} %\hspace{-0.5cm}
  \sum_{0 \leq k_2 \leq k_3 \leq \cdots \leq k_m \leq J}
  \hspace{-0.5cm}
  \Tr
  \big[
    {\cal O}_{M_1}
    Z^{k_2}
    {\cal O}_{M_2}
    Z^{k_3 - k_2}
    \cdots \notag\\
    &\hspace{6cm}
    \cdots
    {\cal O}_{M_{m-1}}
    Z^{k_m - k_{m-1}}
    {\cal O}_{M_m}
    Z^{J - k_m}
  \big]\!(x) 
  e^{ 2 \pi i \sum_{q=2}^mn_q k_q/J}.
\end{align}
Summing over the permutations of
$\{{\cal O}_{M_2},{\cal O}_{M_3},\cdots,{\cal O}_{M_m}\}$, we obtain
the BMN operator given in table \ref{BMN} for a general $m$.

%%%%%%%%%%%%%%%%%%%%%%%%%%%%%%%%%%%%%%%%%%%%%%%%%%%%%%%%%%
%%%%%%%%%%%%%%%%%%%%%%%%%%%%%%%%%%%%%%%%%%%%%%%%%%%%%%%%%%
\section{Beyond the saddle point approximation}
\label{beyond}
%%%%%%%%%%%%%%%%%%%%%%%%%%%%%%%%%%%%%%%%%%%%%%%%%%%%%%%%%%
%%%%%%%%%%%%%%%%%%%%%%%%%%%%%%%%%%%%%%%%%%%%%%%%%%%%%%%%%%
In sec.\ \ref{BMNfromW} and appendix \ref{multi}, we gave analyses
using the saddle point method. 
Thus the validity of the expressions \eqref{saddles2} and \eqref{saddlesm}
is limited only to the region where none of $(k_q - k_{q-1})/J$ nor
$1-k_m/J$ is close to zero. 
However, we can show that this restriction that the
impurities must be well separated can in fact be relaxed;
\eqref{saddlesm} is valid also in the case where two of the $m$
impurities come close to each other.
In this appendix we show this in the case $m=2$.
Generalization to an arbitrary $m$ is straightforward.

Let us consider $F_2(n_2,k_2,J)$ given by \eqref{F_m} with $m=2$ in the
case of $1-k_2/J\sim 0$ (the other case of $k_2/J\sim 0$ can be treated
quite similarly).
Taylor-expanding the phase factor in $F_2$,
$e^{2\pi i n_2\tilde s_2}=\sum_{K}\left(
2\pi i n_2 \tilde s_2\right)^K/K!$, we have
\begin{align}
  F_2(n_2,k_2,J)
  & =
  \sum_{K} \frac{1}{K!} ( 2 \pi i n_2 )^{K}
  \frac{1}{(J-k_2)!k_2!}B(k_2 + K + 1, J - k_2 + 1) ,
  \label{F2exact}
\end{align}
where $B(a,b) = \Gamma(a)\Gamma(b)/\Gamma(a+b)$ is the beta function.
Then we obtain the expansion of $F_2(n_2,k_2,J)$ in powers of $1/J$ as
follows (note that $k_2/J=O(1)$ in the present case):
\begin{align}
  F_2&(n_2,k_2,J)
  = 
  \frac{1}{J J!}\sum_{K=0}^\infty \frac{1}{K!}  (2 \pi i n_2)^K 
  \left( \frac{k_2}{J} \right)^{K}
  \frac{(1+K/k_2)\cdots(1+1/k_2)}
  {(1+(K+1)/J)\cdots(1+1/J)} \notag \\
  &= 
  \frac{1}{J J!}\sum_{K=0}^\infty
  \frac{1}{K!} (2 \pi i n_2)^K \left( \frac{k_2}{J} \right)^K
  \left\{
    1
    + \frac{1}{J}\left(
    \frac{J}{k_2} 
    \frac{K(K+1)}{2}
    - 
    \frac{(K+1)(K+2)}{2}\right)
    +
    O(J^{-2})
  \right\}\notag \\
  &= 
  \frac{1}{J J!} e^{2 \pi i n_2 k_2/J}
  \left\{ 1
  + \frac{1}{J}
  \left( - 2\pi^2n_2^2 \frac{k_2}{J}\left(1-\frac{k_2}{J}\right)
    +
    2 \pi i n_2\left(1 - 2\frac{k_2}{J}\right)
    -1
  \right)
  + O(J^{-2}) \right\}.
  \label{phase2}
\end{align}
The first term in the last line
of \eqref{phase2} agrees with  \eqref{saddlesm} with
$m=2$, and the next $1/J$ term is multiplied by a finite function of
$k_2/J$.
In the other case of $k_2/J\sim 0$, we obtain exactly the same
expression as \eqref{phase2} by starting with the expansion
$e^{2\pi i n_2\tilde s_2}=\sum_{K}\big(
-2\pi i n_2(1-\tilde s_2)\big)^K/K!$.
Eq.\ \eqref{phase2} shows that, so long as $J$ is large, \eqref{BMN2}
is valid even when the two impurities are close to each other.
By the same argument we can validate \eqref{saddlesm} for the case in
which two of the $m$ impurities come close to each other.

%%%%%%%%%%%%%%%%%%%%%%%%%%%
%%%%%%%%%%%%%%%%%%%%%%%%%%%

\end{document}